\documentclass[preprint,12pt]{elsarticle}
\renewcommand{\baselinestretch}{1.1} 

\usepackage{epsfig}
\usepackage{amsmath}
\usepackage{amssymb}
\usepackage{amsthm}
\usepackage{indentfirst}
\usepackage{xspace}
\usepackage{multirow}
\usepackage{hyperref}
\usepackage{xcolor}
\definecolor{darkred}{rgb}{0.2,0.25,0.75}
\hypersetup{colorlinks=true,urlcolor=darkred,linkcolor=darkred,citecolor=darkred}
\usepackage{verbatim}
\usepackage[letterpaper,margin=1.3in]{geometry}
\usepackage{mathpazo}
\usepackage{thmtools}
\usepackage[all]{xy}
\usepackage{longtable}
\usepackage{capt-of}
\usepackage{enumitem}
\usepackage{amsmath,amssymb,graphicx,wasysym,tikz}
 \usepackage[utf8]{inputenc}
\bibpunct[, ]{(}{)}{;}{a}{}{,}% Citation support using natbib.sty
\usepackage[normalem]{ulem}

\definecolor{magenta}{rgb}{1.0, 0.0, 1.0}

\usetikzlibrary{arrows, positioning, shapes, decorations.pathreplacing, decorations.markings, calc, matrix}

\tikzset{->-/.style={decoration={
  markings,
  mark=at position #1 with {\arrow{>}}},postaction={decorate}}}

\usepackage{multicol}

 \usepackage{graphicx}
\usepackage{wrapfig}
\usepackage{lipsum}  % generates filler text

\def\impact#1{\bgroup\narrower%\footnotefont
\baselineskip
\footskip\bigbreak
\hrule\vspace{-0.1 in}\medskip\nobreak\noindent \begin{BI}
\renewcommand{\baselinestretch}{1.1}  {\it #1\/}\par\nobreak\end{BI}}
\def\endimpact{\medskip\nobreak \hrule\bigbreak\egroup}

\setlist{itemsep = 0.20em, topsep = 0.20em}

\declaretheoremstyle[spaceabove=0.25cm,spacebelow=0.25cm,notefont=\normalfont\bfseries, notebraces={(}{)}]{Theorem}
\declaretheoremstyle[spaceabove=0.25cm,spacebelow=0.25cm,bodyfont=\normalfont,notefont=\normalfont\bfseries, notebraces={(}{)}]{noital}
\declaretheoremstyle[spaceabove=0.25cm,spacebelow=0.25cm,bodyfont=\normalfont\color{darkgreen},notefont=\normalfont\bfseries, notebraces={(}{)}]{green}
\declaretheoremstyle[spaceabove=0.25cm,spacebelow=0.25cm,bodyfont=\normalfont,notefont=\normalfont\bfseries,qed=$\qedsymbol$,notebraces={(}{)}]{proofstyle}

\newtheorem{BI}{Broader Impact}[]

\numberwithin{equation}{section}

\usepackage{blindtext}

\usepackage{float}

\begin{document}

\bibliographystyle{utphys}

  \begin{frontmatter}

\title{The Phone Walkers:\\[7pt] \normalsize  A  study of human dependence on inactive  mobile devices}

\author{Laura P. Schaposnik and James Unwin}
 
\address{University of Illinois at Chicago, Chicago, 60647 Illinois, USA.}

\begin{abstract}

The development of mobile phones has largely increased human interactions. Whilst the use of these devices for communication has received significant attention, there has been little analysis of more passive interactions. Through census data on casual social groups, this work suggests a clear pattern of mobile phones being carried in people's hands, without the person using it (that is, not looking at it). Moreover, this study suggests that when individuals join members of the opposite sex there is a clear tendency to stop holding mobile phones whilst walking. Although it is not clear why people hold their phones whilst walking in such large proportions (38$\%$ of solitary women, and 31$\%$ of solitary men), we highlight several possible explanation for holding the device, including the need to advertise status and affluence, to maintain immediate connection with friends and family, and to mitigate feelings related to anxiety and security. 

\smallskip

\end{abstract}


\end{frontmatter}
\smallskip

%%%%%%%%%%%%%%%%%%%%%%%%%%%%%%%%
%%%%%%%%%%%%%%%%%%%%%%%%%%%%%%%%
%%%%%%%%%%%%%%%%%%%%%%%%%%%%%%%%
%%%%%%%%%%%%%%%%%%%%%%%%%%%%%%%%
%%%%%%%%%%%%%%%%%%%%%%%%%%%%%%%%
%%%%%%%%%%%%%%%%%%%%%%%%%%%%%%%%
%%%%%%%%%%%%%%%%%%%%%%%%%%%%%%%%
%%%%%%%%%%%%%%%%%%%%%%%%%%%%%%%%
%%%%%%%%%%%%%%%%%%%%%%%%%%%%%%%%
%%%%%%%%%%%%%%%%%%%%%%%%%%%%%%%%
%%%%%%%%%%%%%%%%%%%%%%%%%%%%%%%%
%%%%%%%%%%%%%%%%%%%%%%%%%%%%%%%%
%%%%%%%%%%%%%%%%%%%%%%%%%%%%%%%%
%%%%%%%%%%%%%%%%%%%%%%%%%%%%%%%%
%%%%%%%%%%%%%%%%%%%%%%%%%%%%%%%%
%%%%%%%%%%%%%%%%%%%%%%%%%%%%%%%%
%%%%%%%%%%%%%%%%%%%%%%%%%%%%%%%%
%%%%%%%%%%%%%%%%%%%%%%%%%%%%%%%%
%%%%%%%%%%%%%%%%%%%%%%%%%%%%%%%%
%%%%%%%%%%%%%%%%%%%%%%%%%%%%%%%%
%%%%%%%%%%%%%%%%%%%%%%%%%%%%%%%%
%\vspace{5mm}
\section{Introduction}

Mobile phones have, without question, been one of the most influential technological developments  in the study of human relations. Since the introduction of phones in 1983, when these devices were only used for most important communications [\cite{history}], to the development of smart phones in 1999 which serve an extraordinary number of purposes [\cite{history2}], phones have played an important role in peoples' lives. As of June 2017, the estimated number of smartphone subscriptions worldwide was 4.6 billion, and growing  at around 4\% per year [\cite{history3}].

\smallbreak

\begin{figure}[H]
\centering
\resizebox*{6.5cm}{!}{\includegraphics{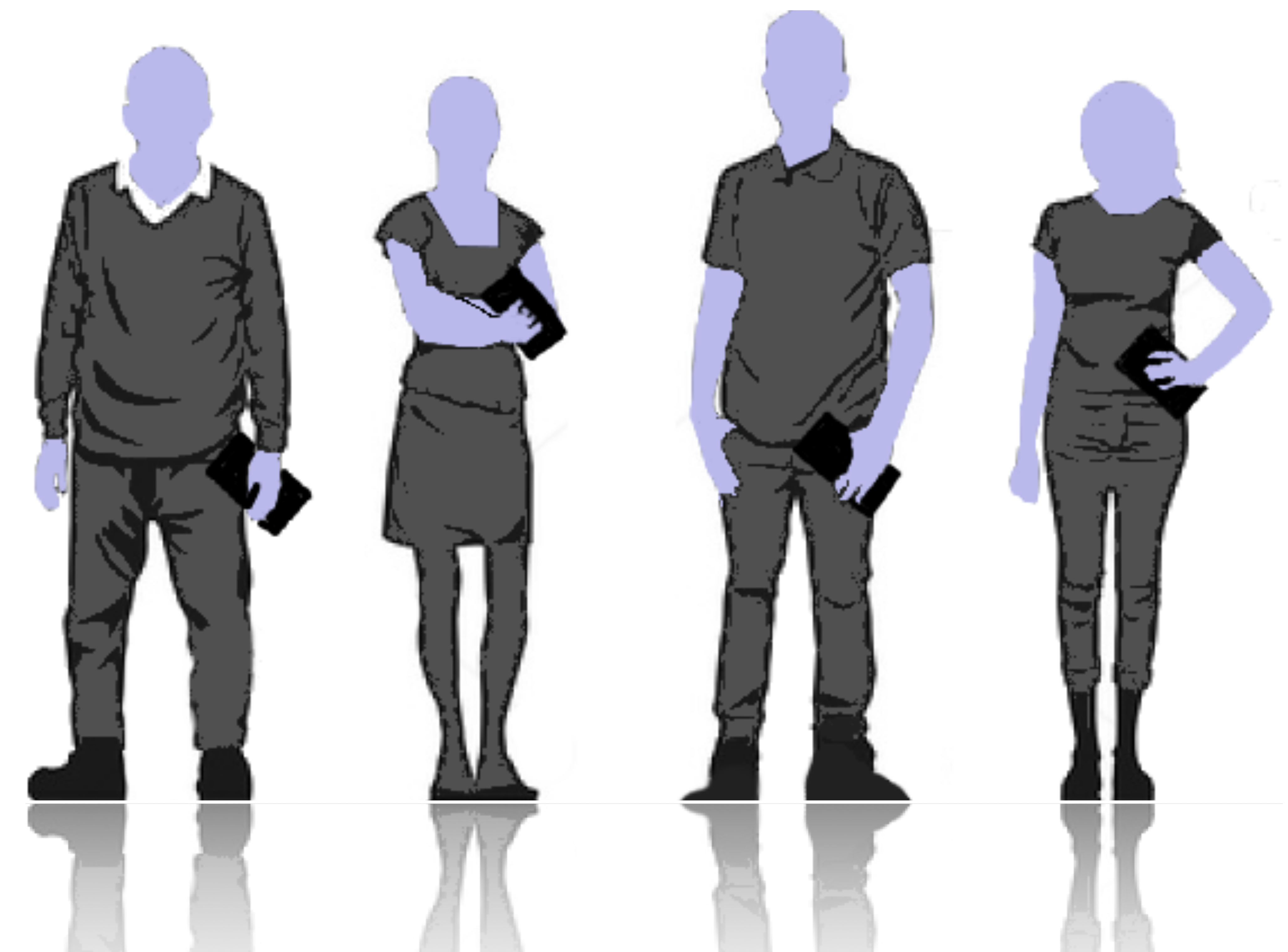}}
\caption{Four of the most common poses of humans holding phones whilst walking.}\label{Figure0}
\end{figure}

 The effect of mobile phones on human relationships, and the different ways in which  they are used for communication, have been topics of much study in recent years (see, for instance, \cite{dunbar1}). The present paper reports on a previously unobserved phenomena of people visibly holding their phones for long durations without using the device.
We refer to these as {\it phone walkers} and common poses are illustrated in Figure \ref{Figure0}.

\vspace{-2mm}
\vspace{-2mm}

 Through the present study,  we found that surprisingly a very high proportion of pedestrians were  phone walkers and the percentages increased when considered only females.
  Moreover, as it shall be seen, people change their attitude towards holding mobile devices when paired with people of their same sex or, especially, the opposite sex. Thus, it is clearly of interest to understand the changes in    peoples' behaviour  towards, and dependence on, inactive mobile devices in relation to their genders: this is what   the present manuscript is dedicated to.

At present there is not one clear explanation as to why phone walking is so prevalent among humans, and in particular why this is more pronounced in the female population observed.   However,  note that object carrying as socio-sexual display has been studied in the context of other mammals (see, for example, [\cite{AGUA}, \cite{rat}]), and also within human infants [\cite{baby}]. The study of phone walkers provides a good setting from where to consider the past literature on object carrying in the context of the present day technology. 
Moreover, the apparent predisposition of significant swathes of the population to maintain continuous contact with their mobile devices is notable and  insightful parallels might be drawn with the well established research  studies of mobile phone addiction  [\cite{addict,addict-gender2,addict-gender}]. Indeed, the growing reliance and addictive tendencies in connection to technology has been identified as a emerging issue of concern and, in particular,  online gaming addiction [\cite{gaming1,gaming2}] has already been labelled as a ``Condition for Further Study'' in the DSM-5 [\cite{DSM}]. 
  
Notably, studies of both mobile phone and online gaming addiction have indicated that gender may play a leading role in determining the likelihood of an individual becoming reliant or addicted to a given form of technology. Specifically, it has been suggested that females are significantly more likely to develop an addictive relationship with mobile phones, whilst males are more likely to experience a similar reaction with online gaming  [\cite{usage-gender,usage-gender2}]. Indeed, studies have found that females demonstrate higher levels of problematic phone use, see for instance [\cite{addict}] and references therein.  It has also been shown that individuals' susceptibility to becoming reliant or addicted to mobile or other technology may also be influenced by their nationality or ethnic origins [\cite{ethnicity1,ethnicity2}], and socioeconomic class [\cite{class3,ethnicity3}]. However, differences between these subgroups appear less pronounced,  and the conclusions less robust than the observed impact of gender. Because of this, in the present work we will primarily focus on differences in behavior in the adult population due to the different gender compositions of groups of individuals or pairs. In the Discussion section we will return to examine  the most likely explanations for phone walking in humans and  in the Conclusions the possible implications of existing studies of active phone usage between individuals of differing age, ethnicity, and class  is discussed in the context of phone walkers.

  This is the first work in a series of papers which will explore the relationship between people and their mobile devices.   This first study we recorded only the gender and group size of our subjects and thus we focus on gender differences in the total adult population. Remarkably, this already reveals some fascinating behavioral differences which we shall describe throughout the paper. Moreover, studies which  divide the population according to age, class, and ethnicity are interesting and important, and shall be returned to in subsequent work, as will be discussed in the conclusions section.

  \section{Methods}

To explore the above questions, a large number of casually walking groups (of singletons and pairs) were sampled in natural social settings in the centre of a European capital, and their use of mobile phones was analysed. These data samples were used to examine the size and sex composition of social groups, as well as the presence of mobile devices whilst walking, both for communication or simply being visibly held without being used.
    
    \newpage
    %%%%%%
 A total of 2209 casually walking social groups were sampled  at several locations in central Paris and away from major tourist sites. The sample involved 3038 adults (1633 females; aged 21-65, with the  estimated mean age 35 years). The venues included 6 different districts in Paris, and locations were  chosen such that they exhibited a high rate of pedestrian traffic. On arrival at a location, all walking social groups were noted and observed for 20-30 meters.\footnote{The street width in Paris is typically 10-20 meters (see e.g.~\cite{paris}) and thus a range of 20-30 meters corresponds  to the typical line of sight available for stationary observations. This provided an adequate period to obtain an accurate spontaneous observation.}  All sampling was carried out by LS and JU at a variety of hours of the day between 11.00 -18.00, on both working days and weekend days, and the data is presented in Table A.1. and Table A.2. of the Appendix. 

 A walking social group was defined as a set of people who were clearly with each other (walking together or talking to each other), and a  phone walker was defined as a person who, without using their phone, was carrying the device in their hand (see Figure \ref{Figure0}).   In the vast majority, the groups encountered were composed of at most 2 people.  For each social group observed, only the  following were noted: total number of people in the walking group (group size being 1, 2, or $>2$), the apparent sex of each individual (male or female), and whether individuals were visibly holding a phone.  
 
 When a person holding a phone was observed, it was noted whether    the subject was interacting with the device or not. Throughout the paper,  a person of the latter type shall be referred to as   a {\it phone walker}.   Groups of size 2 were recorded and classified as single sex female (two females), single sex male (two males), or mixed sex (one male and one female). Observations were passive, and there were no interventions or interactions with the subjects.  In particular, no private or personal identifiable information was  recorded. 
 
 Apparent sex was identified based on the assumption that subjects would conform to standard cultural norms. It is expected that  the rate of misidentification of sex is negligible. Identification of phone holding was performed by an observation of both hands of each subject, and it is also expected that  the  misidentification rate is negligible.  Observations were made spontaneously (over 20-30 meters) and thus were not revised should a subject subsequently extract or store their phone after recording.  
 
 The inclusion-exclusion criteria for population sampling was chosen in order to obtain overall estimates for the adult population, and thus only groups in which all individuals were between 21 and 65 years of age were recorded.  Apparent age of individuals was assessed via a set of common visual cues (e.g. see \cite{Rhodes}), through observations of height, build, and facial features, following similar practices to  \cite{Robin,Robin2}.  In particular,  the number of individuals determined to be over 65 years accounted for less than 1\% of cases and fewer than 2\% of cases were determined to be under 21 years, the vast majority of which were infants. It is expected that the rate of misidentification of age eligibility is negligible. Notably, adolescents (similarly the elderly) may use their phones differently to the overall adult population, and hence the sampling was restricted to the adult group of ages 21-65 in order to obtain robust conclusions. 
  
As discussed further in the conclusion, it would indeed be most interesting for future work to analyse the results of the present study in the context of smaller ranges of ages, in particular to understand the different behaviours that older generations have when compared with teenagers. Finally, it should be also mentioned that whilst we restricted our observation to groups of size 1 or 2, the number of groups of bigger sizes were negligible (lower than 1\%).

\vspace{-1mm}
   \section{Results}
\vspace{-1mm}

Passive observations were made of the 3038 adults, of which 674 (396 females) were identified as phone walkers, accounting for 22.1\% of the total population.  Most interestingly, whilst 19.7\% of the male population observed were phone walkers, the rate of phone walking increased by more than half, to 33.3\% when considering only females, suggesting an important difference between sexes when studying phone walkers.  In all instances observed, the phone walkers had bags or suitably large pockets in which  devices could be stored if desired, thus it is understood that  individuals were not carrying their  phones in their hand out of necessity.

The people observed were noted as   singletons as well as same-sex and mixed-sex pairs, and in the case of pairs we recorded instances in which one or both members of the couple were phone walkers. The group breakdown of the 3038 people in the data set was given by 1380 solitary walkers (730 females), 310 single sex pairs (192 of females), and 519 mixed sex pairs. For single sex groups, the mean group sizes observed were 1.15 for the male-only groups (N = 768), and 1.20 for the female-only groups (N = 922). 
The dataset was recorded in 8 subsets taken at 6 locations. The total number of each group observed and the percentage of groups with one or more phone walkers is reported in Table 1, along with the standard error from the mean percentage across the 8 datasets. 

%\smallskip

\begin{table}[h]
\begin{center}
\begin{tabular}{|c|c|c|c|c|c|c|c|}
\hline
Group & Total Number & \% Phone Walking & Standard Error \\
\hline
All persons & 3038  &  22.1  & 2.6 \\
\hline
Single Male & 650 & 30.4 &  1.3\\
Single Female & 730 & 37.1  & 1.9\\
\hline
Two Males & 118 & 24.4  & 2.4\\
Two Females & 192 & 40.3  & 5.3\\
Two mixed & 519 & 18.4   & 3.2\\
\hline
All Males & 1405 &19.7 & 0.6\\
All Females &1633  &33.3  & 2.7\\
\hline
\end{tabular}
\caption{Summary of observations showing the total number of each group observed and the percentage of groups with one or more phone walkers. See Appendix for further tables of observations.
 \label{table1}}
\end{center}
\vspace{-5mm}
\end{table}
\pagebreak

Naively, one might expect only a small fraction of individuals to be carrying inactive phones on display, corresponding to those yet to store their devices or people intending to use them imminently. We can test whether a sizeable or diminutive fraction of the population carry inactive devices by constructing a null hypothesis $H_0$ which assumes that some (arbitrary) non-negligible fraction of the population are phone walkers. Specifically, we compare our data set against a null hypothesis in which 5\% of the population are assumed to be phone walkers, or in other words, that the probability of a given individual being a phone walker is $P_0 = 0.05$. If the observed fraction $P$ of phone walkers is substantially greater than $P_0$, then this will be indicated by a $\chi^2$ value for  $P$ which is significantly smaller than the value calculated for $P_0$. By inspection of Table 1 one can intuitively infer that our dataset suggests quite different behavior from this null hypothesis. Our complete set of observations, including the 8 individual data subsets, is given in the Appendix. 

 The refutation of the null hypothesis can be quantified via a parametric test, by considering $P$ the proportion of phone walkers within the observed population, and letting $P_0:=0.05$ be the proportion assumed in the null hypothesis. To perform the parametric test, consider the following null hypothesis $H_0$ and alternative hypotheses $H_\pm$ defined as:

 \hspace{1mm}

$H_0:$~``$P=P_0$,  phone walkers comprise \textit{\textbf{exactly}} 5\% of the population.''

 \hspace{1mm}
 
$H_+$:~``$P> P_0$,  phone walkers comprise \textit{\textbf{more}} than 5\% of the population.''

 \hspace{1mm}
 
$H_-$:~``$P< P_0$,  phone walkers comprise \textit{\textbf{less}} than 5\% of the population.''

 \hspace{1mm}
 
Denoting by $P_i$ the proportion of phone walkers in each data set of $N_i$ individuals, then the $\chi^2$ for the null hypothesis with $P_0=0.01$ is found to be 
\begin{equation*}
\chi^2(H_0)=\frac{1}{P_0}\sum_{i=1}^8(P_i-P_0)^2N_i\approx 2048,
\end{equation*}
where the sum is made over the eight independent datasets, as presented in the Appendix, and thus the $\chi^2$ test is given relative to 7 degrees of freedom.
From the above, one can see that the null hypothesis $H_0$ is excluded with strong significance (with a $p$-factor  below 0.0001). The alternative hypothesis $H_-$ with $P< P_0$ implies $\chi^2\gg2048$ is even more strongly disfavored and thus the validated alternative hypothesis is $H_+$ with $P> P_0$ and $\chi^2<2048$.  Therefore we conclude that the data strongly suggests that the rate of phone walking is above 5\% of the population, or equivalently,   that:

\vspace{2mm}
\indent {\em Phone walking is observed in a significant fraction of the population.}
\vspace{2mm}

\begin{figure}[H]
\centering
%\vspace{10mm}
\resizebox*{11cm}{!}{\includegraphics{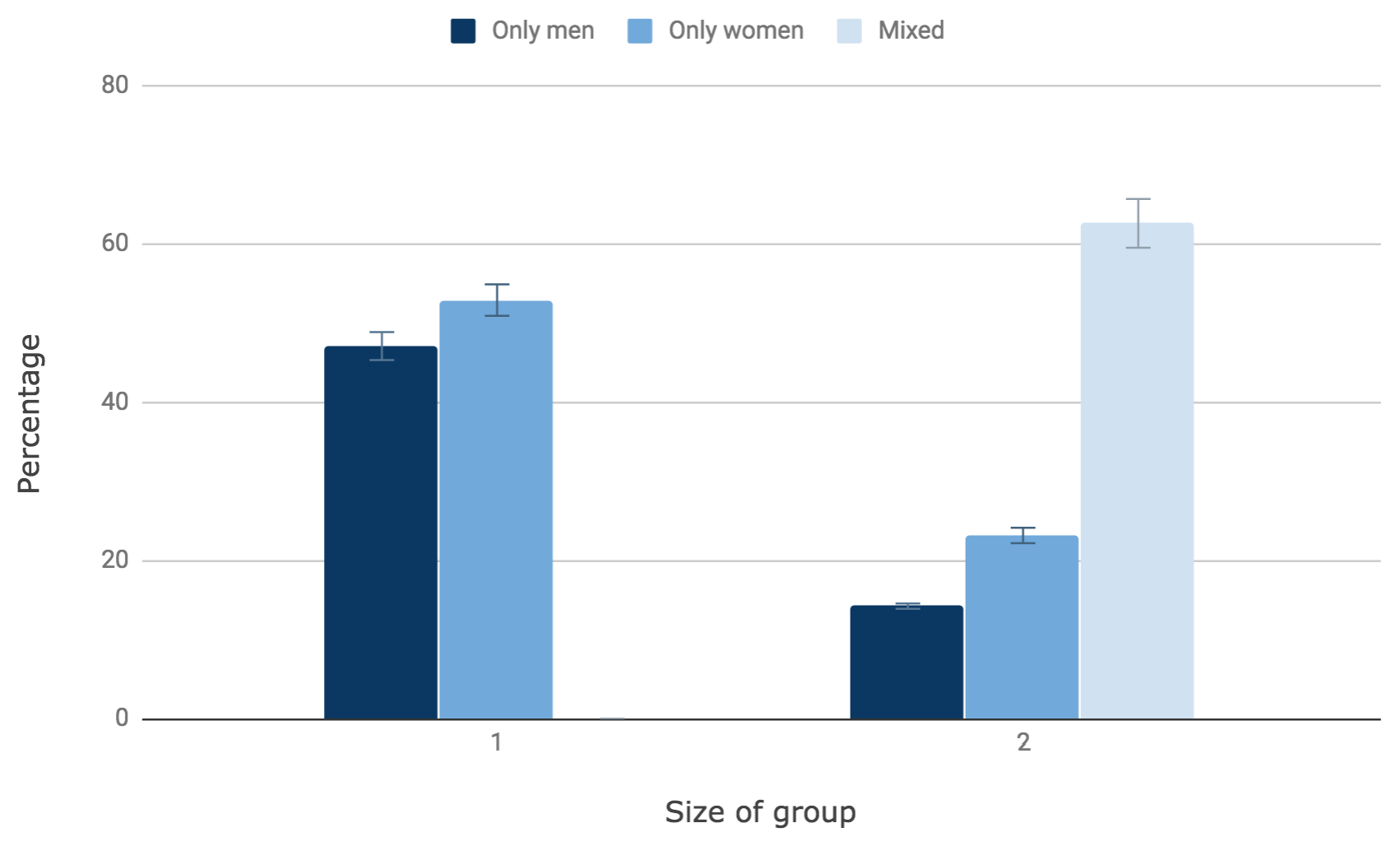}}
\caption{Distribution of social group sizes for male only ($N = 768$), female only ($N = 922$), and mixed sex ($N = 519$) groups. Standard errors as seen in Table C.1. in the Appendix. }\label{Figure1}
\end{figure}

Much can be learned from the dataset acquired by considering the breakdowns with respect to group size and gender. The distribution of walking group sizes for single sex and mixed sex social groups is illustrated in Figure \ref{Figure1}. Overall, the  social groups encountered had a size of 1.37 adults (range 1-2), and the mean number of phone walkers per group was 0.30 (range 0-2). No group larger than 2 people was  included in the analysis,  since these constituted less than 0.05$\%$  of the groups observed.

Whilst the percentage of male and female walkers   were almost equivalent (53.8$\%$ of women),  one can see from the   Figure  \ref{Figure1} that pairs of people were almost three times more likely to be  of mixed sex than only women (2.7 mixed sex pairs for each 1 pair of women). Moreover, pairs were four times more likely to be of mixed sex than only men (4.4 mixed sex pair for each 1 pair of men).  Given these likelihoods, it is very interesting that the number of phone walkers within mixed sex pairs dropped considerably compared to all other groups, as   shall  be seen in the following sections. 

\begin{figure}[H]
\centering
\resizebox*{14cm}{!}{\includegraphics{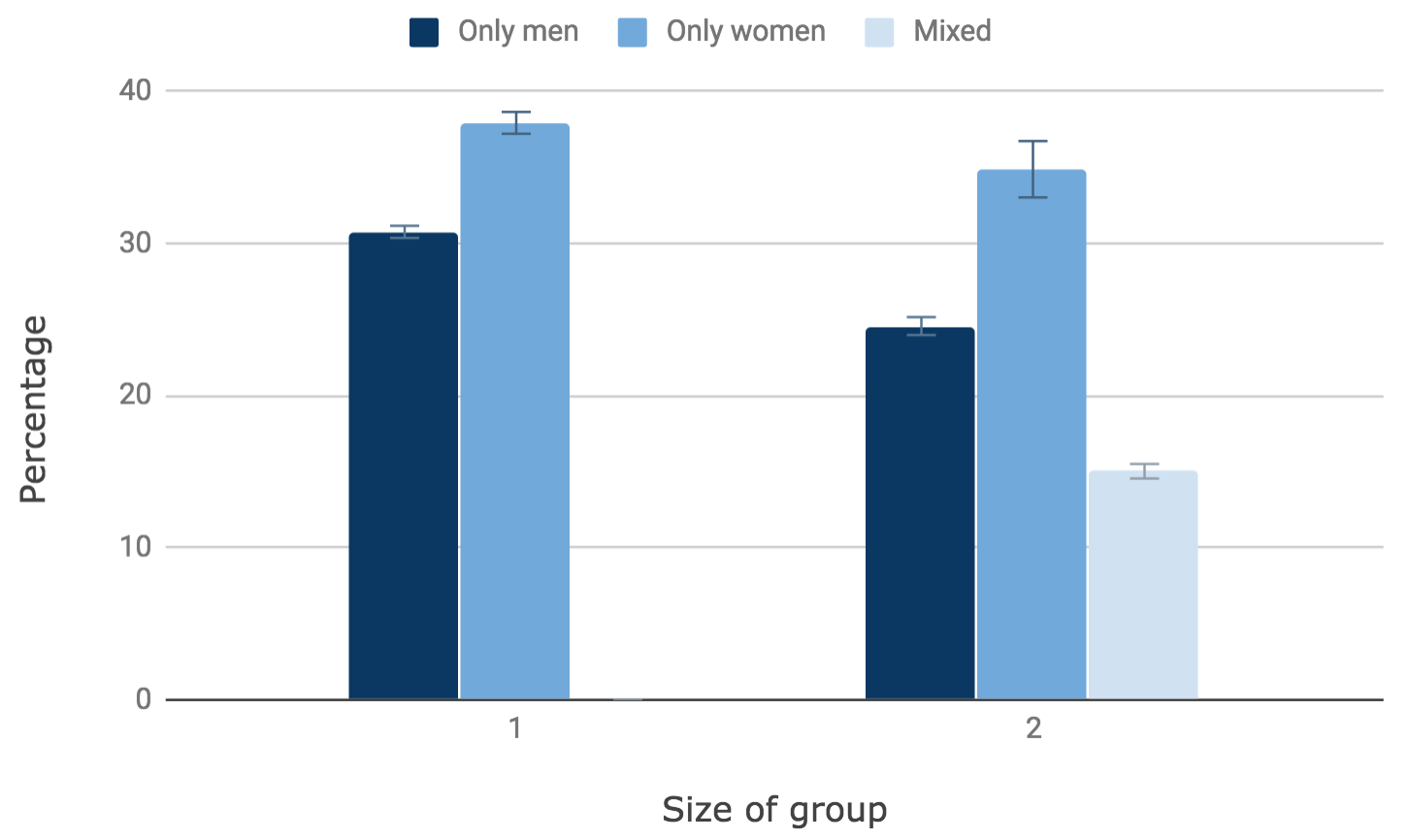}}
\caption{Distribution of groups with at least one phone walker for male only ($N = 768$), female only ($N = 922$) and mixed gender ($N = 519$) groups. Standard error as seen in Table 1 above. }\label{Figure2}
\end{figure}

In order to understand phone walking, Figure \ref{Figure2} plots the percentage of groups with at least one phone walker, against the  group size for mixed sex,  and the two single sex social groups.  In the case of the 1380 solitary people (730 females),  37.9$\%$ of females and 34.9$\%$ of males were  phone walkers. When considering all observed pairs of walkers, 
  whilst mixed sex groups were the most common among all pairs (519 of the total observed of $N=829$),   the percentage of mixed sex groups that had at least one phone walker was the smallest. Indeed, only 15$\%$ of all mixed sex pairs had a phone walker, whereas for pairs of women  this percentage  was twice as large.

\vspace{10mm}

\begin{figure}[H]
\centering
\resizebox*{12 cm}{!}{\includegraphics{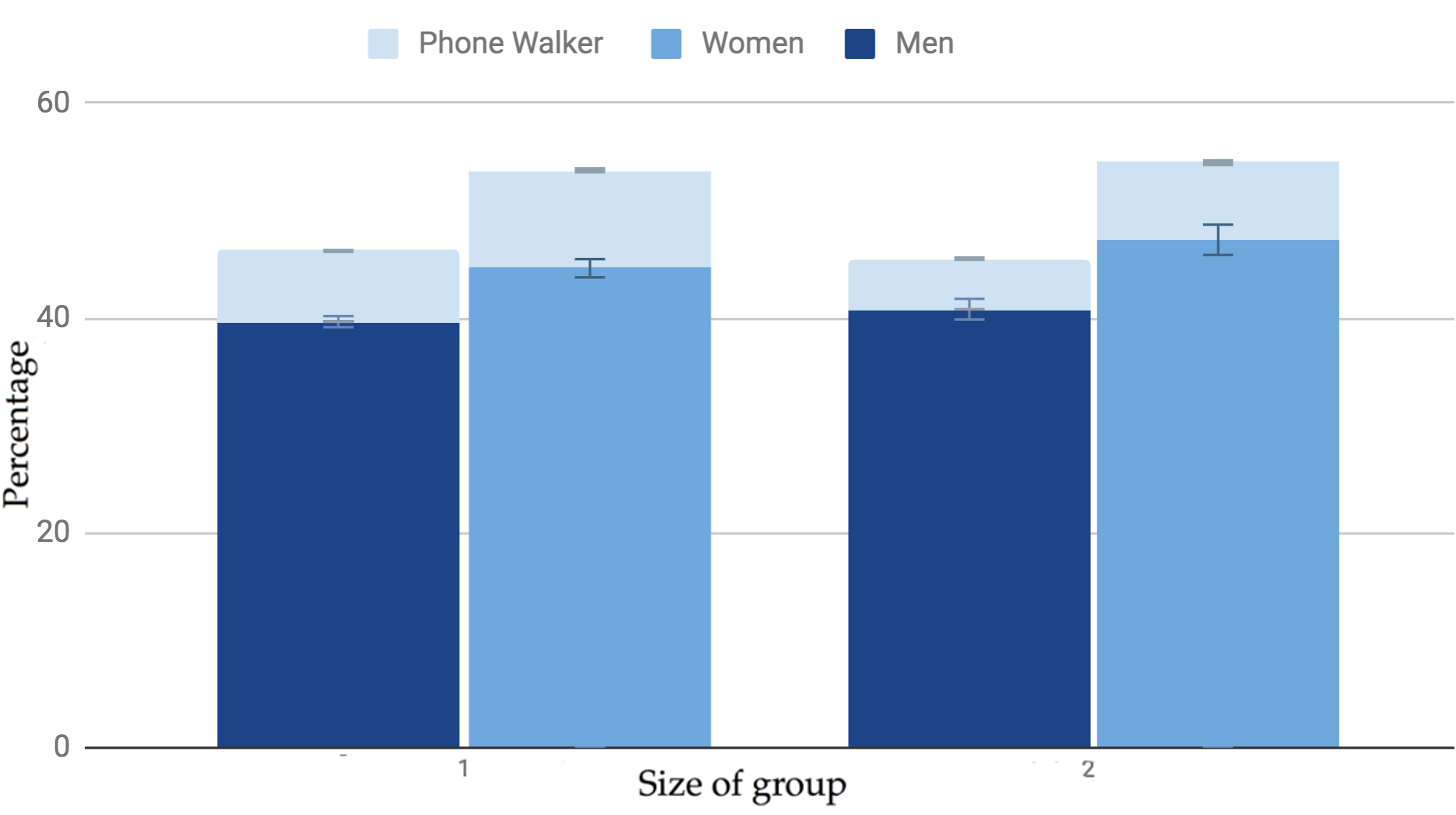}}
\caption{Percentage of phone walkers of each sex in groups of each size (total of $N=2209$ groups).  Standard errors as appearing in  Table C.3. and Table C.4. of the Appendix. }\label{Figure4} 
\end{figure}

  In order to highlight the relevance of gender within this study, we considered phone walkers of each gender, in groups of each size.  In Figure \ref{Figure4} one can see the percentage of phone users of each sex in groups of each size, where groups of 2 people have been considered together, whether they are of mixed sex or not.  
  In particular, it should be noted that whilst in  7.8$\%$ of female pairs  both women were phone walkers, this percentage halved when considering male pairs for which  both people were  phone walkers (sample size of 192 pairs of women, and 118 pairs of men).

\begin{figure}[H]
\centering
\resizebox*{10cm}{!}{\includegraphics{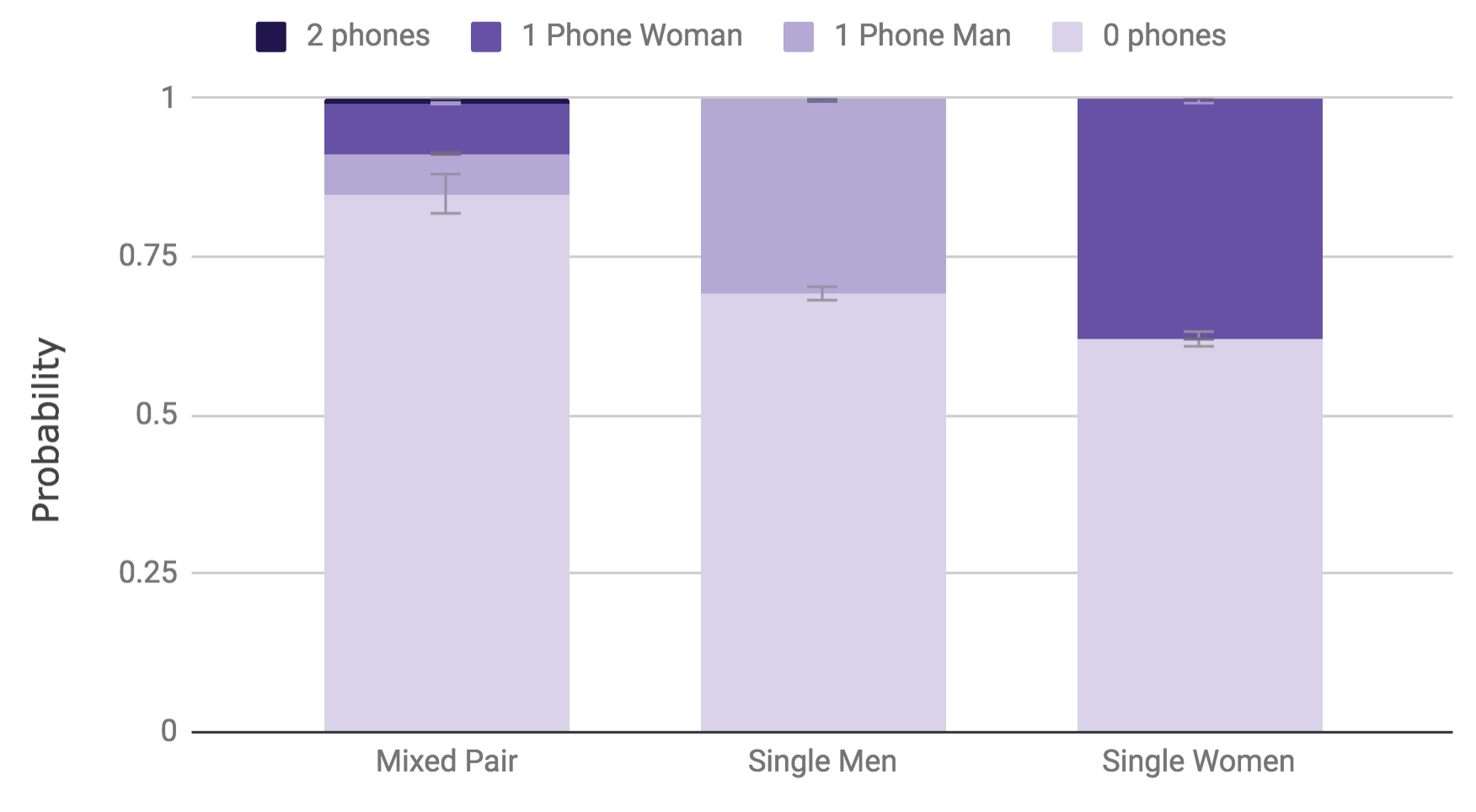}}
\caption{Male/female solitary phone walkers vs phone walkers in mixed sex pairs. Standard errors as appearing in Table C.2. and Table C.3. of the Appendix.}\label{Figure6}
\end{figure}

The number of solitary phone walkers (both male and female) can be compared to the appearance of phone walkers within pairs of people, as done in  Figure \ref{Figure6}. In particular, one should note that although 37.9$\%$ of solitary women were phone walkers (30.8$\%$ of men), these percentages dropped drastically when men and women were seen in a mixed sex pair. Indeed, out of all mixed sex pairs, the percentage for which only the woman was a phone walker was 7.9$\%$ (and  6.4$\%$ for men).  This is, of all mixed sex pairs with exactly one phone walker,  55.2\% of them comprised a female phone walker. Therefore, one can see that in a mixed sex pair with exactly one phone walker, the walker is 20\% more likely to be a  woman.
Finally,  less than 1$\%$ of people in a mixed sex pair were both phone walkers, indicating a remarkable decrease in phone walkers when joined by a person of the opposite sex -  this shall be further studied  in Figure \ref{Figure7}.  

\begin{figure}[H]
\centering
\resizebox*{10cm}{!}{\includegraphics{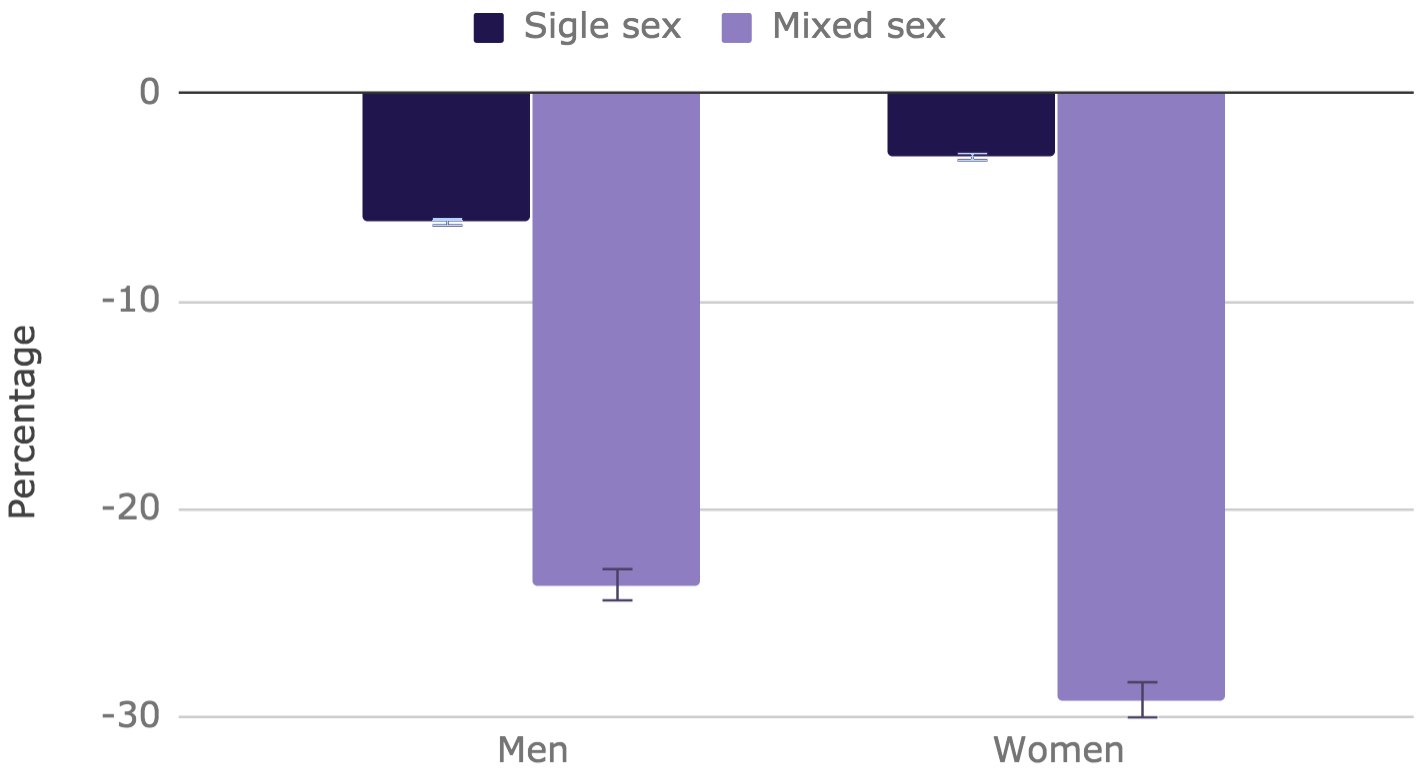}}
\caption{Percentage drop of phone walkers in pairs with at least one phone walker compared to solitary males and females. Of the total number of groups observed, the number of groups with phone walkers was of $N = 229$ for male only, $N = 344$ for female only, and $N = 78$ for mixed sex. Standard errors as appearing in Table C.2. and Table C.3. of the Appendix. }\label{Figure3}
\end{figure}

\begin{figure}[H]
\centering
\resizebox*{12cm}{!}{\includegraphics{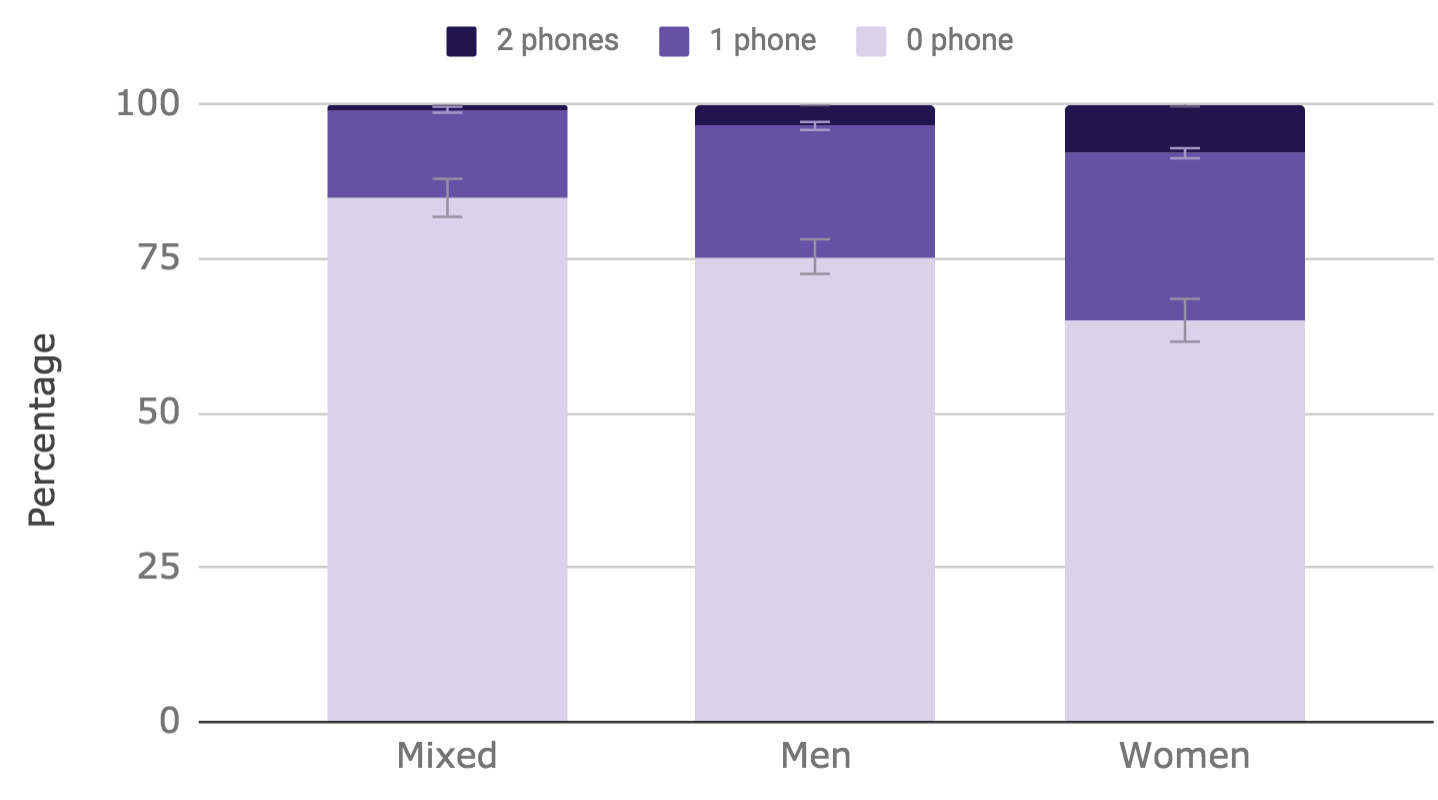}}
\caption{Phone walkers in all three types of pairs, where the total of pairs observed is $N = 829$, of which 310 were single sex pairs.  Standard error as appearing in Table C.2. of the Appendix.}\label{Figure7}
\end{figure}

As mentioned above, it is very interesting to note the the percentage drop of the number of groups which have a phone walker, depending on whether the social groups were single sex or mixed, which is plotted in Figure \ref{Figure3}. In particular, note that although the rate of female phone walking did not drop significantly for groups of two women compared to solitary females, the drop in phone walking is considerable when accompanied by a walker of the opposite sex.  
Specifically, compared to solitary females (or males),  whilst a drop of 6.2$\%$ in the number of cases of phone walkers for pairs of women was observed (3.0$\%$ for pairs of men),   a drop of 29.3$\%$ was seen for female phone walkers accompanied by a male (the drop was of 23.6$\%$ for a male phone walker joined by a woman).

%\newpage

To understand how single sex pairs behave, Figure \ref{Figure7} shows the percentage of different types of phone walkers within pairs of people. In particular, whilst only   3.4$\%$ of pairs of men involved two phone walkers, the percentage doubled for female pairs. In the case of only one phone walker within a single sex pair, one could notice that the female pairs were 27.8$\%$ more likely to have a phone walker than male pairs. Indeed, 21.2$\%$ of pairs of men had exactly one phone walker, whilst 27.1$\%$ of pairs of women had exactly one phone walker. 
 Note that the percentages of pairs  with one phone walker decrease considerably from female pairs, to male pairs, to mixed sex pairs.

The observations in Figure \ref{Figure3} and Figure \ref{Figure7} inform our second main hypothesis:

\vspace{2mm}
\begin{center}
{\em ``Phone walking is significantly less prevalent in mixed sex pairs than singletons or same sex pairs''.}
\end{center}
\vspace{2mm}

We assess the above hypothesis within our dataset by implementing a $\chi^2$ test, specifically we test the following null hypothesis $H_0$, and alternative hypothesis $H_a$ by considering $P^*_i$  the proportion of phone walkers for same sex pairs in dataset $i$, and letting $P_i$ be the proportion of phone walkers for mixed sex pairs in dataset $i$. To perform the parametric test, consider:

 \hspace{1mm}

$H_0:~P^*_i=P_i$, within pairs of walkers, {\bf there is  no association} between gender  composition and the appearance of phone walkers (the percentage of phone walkers is the
same for  same sex pairs  and mixed sex pairs).
 \hspace{1mm}

$H_a:~ P^*_i>P_i$, within pairs of walkers, {\bf there is an association} between gender composition  and the appearance of phone walkers (the percentage of phone walkers in same sex pairs is greater than the one in mixed sex pairs).

 \hspace{1mm}

The the null hypothesis $H_0$ gives a $\chi^2=54$ on 14 degrees of freedom and thus is excluded with strong significance, with a $p$-factor below 0.0001 and therefore the alternative hypothesis $H_a$ is strongly favoured. Furthermore, note that with 14 degrees of freedom $\chi^2\geq30$ implies a statistically significant result, with $p<0.01$. The full datasets are presented in the Appendix, and a similar hypothesis test can be made for singletons, in which case we find  similarly that the null hypothesis {\em ``that the rate of phone walkings in singletons and mixed sex pairs are equal''} is also strongly disfavoured, with a $\chi^2\approx59$ on 14 degrees of freedom. Hence, in this case the following alternative hypothesis  is well  supported by the data:

 \begin{center}
{\em ``Phone walking is  less prevalent in mixed sex pairs than singletons''.}
\end{center}

 \section{Discussion}

The phenomenon of  phone walking identified in the previous sections may  arise for different reasons  depending on the sex of the individual. However, considering the   advancement of technology, and the approach that humans have to mobile communications, one should also consider broader reasons for the existence of  phone walkers of both sexes. As noted previously individuals were not carrying their phones out of necessity, having bags and pockets suitable to store the device.  
In what follows four global reasons for men and women to become phone walkers are proposed:
{\it immediate availability};  {\it psychological dependency}; {\it personal safety};  and {\it social appearance.}
\smallbreak
\smallbreak

 \noindent \textit{\textbf{Immediate availability.}} As mentioned in innumerable news articles, 
{\it  ``Text messages magically appear on the screen of your phone and demand immediate attention from you.''} [\cite{Guardian}]. It is therefore expected that, if one wants to be part of the  constantly evolving conversation, mobile devices should be always ready to be used instantaneously. Moreover, as mentioned in [\cite{attention}], there is a tendency for people to anticipate the availability of a person before making a call or sending a message.  Hence,  there is  a common need among people to make it clear to themselves, as well as to those observing them, that they are indeed available and ready to receive incoming communication. 
 
 \smallbreak

The notion of immediacy within social media is most important when studying phone walkers.  At a time when the industry is continuously  encouraging users to be constantly checking, updating, and commenting on the content posted by other users, 
 the result is a culture which praises general availability (and not only  for communications), and in which a large percentage of users  measure their success and happiness in terms of    their standing in social media.  Indeed,  several studies have  related  online socialising and information exchange to personality [\cite{tale}].  
   Users build their relationships via their reactions on social media (which can largely overlap with social standing in reality). In doing so,   users are judged on the intensity and expediency of their reactions to the posts of others. Reciprocal validation permeates social media and there are implicit expectations, fostered by the industry, that one should expect a similar level of online attention as one is willing to commit to the posts of others, hence making apparent the need for walkers to always have their phones at their fingertips.

\smallbreak

It has been noted in the present study that the rate of  phone walkers in mixed sex pairs is dramatically lower than for solitary walkers, where the percentage drop compared to solitary female walkers is around 30\%.  Studies   have shown that romantic partners expect text messages to be answered within 5 minutes (for instance \cite{touch}),  thus one may indeed expect the number of phone walkers to drop within pairs of people of opposite sex, considering the proportion of heterosexual partnerships in Europe.    Moreover, since some fraction of the same sex pairs observed  are  also most likely  in committed romantic relationships,    this could also explain the drop in phone walking in single sex pairs. 
   
   \smallbreak

   One of the principal biological aims of building a social network  is to facilitate the finding of a suitable mate. In particular,  the size of a social network has been shown to impact the success in finding mating partners, both for humans as well as for other animals: see, for example, the study of long-tailed manakin in  [\cite{tail}]. Ultimately, after a stable mating pair is formed,   same sex interactions are often neglected in favour of building a stronger bond with  a romantic partner. Indeed, it has been shown that whilst at the start of a romantic relationship the development of new common friends is favoured, when the couple transitions to living under the same roof, the overall level of sociability drops  [\cite{drops}]. Thus the observation that phone walking is less prevalent among mixed sex pairs might be a specific example of a broader neglect of other relationships for people when in stable romantic relationships.

  \smallbreak

      From another perspective, one should recall that there where notable differences between male and female  phone walkers and their change in attitudes when joined by other people whilst walking.  Interestingly, when considering availability of people for communication, and studying parallel communication habits, it has been shown in  [\cite{muller}] that no sex differences could be observed. Hence, whilst generic availability may be a reasonable explanation for  phone walkers to appear, the difference in sex requires further analysis and we discuss possible implications of gender differences in greater detail in the conclusions.

  \smallbreak

 \noindent \textit{\textbf{Psychological dependency.}}  The pressure on individuals to be immediately availability can lead to an increased use of mobile devices, which in turn can result in anxiety if separated from the mobile device. Thus it is quite conceivable that the simple manipulation of the object could lead to a corresponding decrease in tension or anxiety compared to when the phone is stored in a bag or pocket. This form of psychological dependency has been studied   extensively in connection to technology addiction, see for example [\cite{addict,addict-gender2,addict-gender}].
 
 \smallbreak

The results in the present paper indicate that phone walking behaviour is more prevalent in females than males, and it is interesting to compare with gender differences reported in the context of mobile phone addiction. Notably, it has been observed that women are more likely to   use mobile phones and more likely to develop dependencies [\cite{Billieux,Walsh10,Walsh11}].

   \smallbreak
 \noindent \textit{\textbf{Personal safety.}}  In recent years many researchers have considered different ways in which mobile phones have influenced people's life (see, for instance [\cite{wei}] and [\cite{safe}]). From a relevant perspective, when studying the impact of mobile devices on young adults, it was shown that mobile devices  {\it``are ameliorating fear of crime and other dangers associated with being outdoors, and helping young people to self-empower in their use of public spaces''} [\cite{safe}]. 
Then,   it seems quite plausible that individuals may hold their phones both for personal reassurance against perceived threats, and as a visible warning sign to potential assailants. It is also worth noting that holding the phone may reduce the risk of larceny by pickpocketing.
  
 \smallbreak
\smallbreak

 \noindent \textit{\textbf{Social appearance.}} Humans  have long been carrying accessories and marks to display their social standing within their population [\cite{material}]. Smart phones are typically highly valuable objects [\cite{apple}], and  thus by visibly displaying smart phone devices prominently,  an individual can advertise their affluence and social standing. This is not incomparable to displays of affluence by wearing designer fashion clothes or jewellery: ultimately, the aim of such displays is typically thought to be to enhance or affirm a person's social standing and to attract a suitable mate. Thus, by carrying a mobile device visible to the observers, even when not in use, humans are displaying their social status.

%%%%%%%
 
%%%%%%%

\section{Summary and Conclusions}

The present paper reports on a previously unidentified phenomenon of humans visibly holding mobile devices for moderate durations without using them, which is referred to as  {\em phone walking}. A large data set of humans walking in central streets of a European capital was gathered though spontaneous observations. Using the data set of 2209 casually walking social groups, composed of 3038 individuals, it was found that a significant fraction exhibited this behaviour (see Figure \ref{Figure4}). A modest gender bias was observed, with this phone walking behaviour being more prevalent among women than men. Moreover, it has been highlighted that there is a substantial reduction in phone walking for humans in mixed sex pairs (see Figure \ref{Figure3}). It is proposed that   phone walking could arise due to a combination of social pressure to be available,  security concerns, physiological dependance and exhibition of status symbols.

The tendency for individuals to be phone walkers was seen to drop drastically when accompanied by another individual,  especially  in the case of a member of the opposite sex. The theories regarding this drop in phone walking within mixed sex couples can be understood if the majority of mixed sex pairs commonly constitute romantic couples, and studies in the literature provide evidence to suggest that mixed sex pairs in public settings are commonly in romantic relationships  [\cite{Adams}]. Some fraction of the same sex pairs are likely also in romantic relationships, and this may explain the mild drop in phone walking in single sex pairs compared to individuals. However, it is difficult to estimate the number of same sex couples from the  data acquired through passive observation.

The data for this study was gathered at several locations in central Paris. Whilst it seems quite reasonable to propose that Paris may be representative of Western European culture, it would be desirable to verify this with further studies of other cities and municipalities. Moreover, it is not necessarily apparent that this behaviour should be inherent in the population of other parts of the world,  and thus   wider studies would be of interest. In particular, one might expect that the relationship between humans and their mobile devices has evolved overtime, and changed as the sophistication of mobile devices has developed. For a general overview of mobile phone penetration in developing countries, see for instance \cite{Donner}.  Thus, a similar study in societies with less exposure to cell phones, such as Africa or South America, would be of interest in particular since different socioeconomic classes have been shown to have different levels of ownership and reliance on mobile phone technology [\cite{class1,class2}]. It could also be of interest to repeat our study in highly technologically integrated societies,  such as South Korea, where research has suggested that mobile phone addiction is especially high, being at twice the rate of the United States (US) [\cite{85}].

Before closing, it is interesting to consider how the rate of phone walking may differ between population subdivsions other than by gender. In particular, age, ethnicity, and class have each been studied to a significant degree in the literature on mobile phone addiction and thus implications for phone walking might be inferred by drawing analogies. Certain studies have indicated that mobile phone addiction may be more prevalent in minority populations, for instance \cite{ethnicity2}. However, it is often difficult to disentangle ethnicity from socioeconomic class (especially, for instance, in the US). Notably, it has been suggested in \cite{ethnicity1} that minorities in the US are typically more reliant on their phones for online access, with 13\% of Hispanics and 12\% of African Americans  entirely dependent on their smartphones for internet access in comparison to 4\% for Americans of European descent. 

Although existing studies have found very strong correlation to age and gender in phone usage and addiction, studies of social class, ethnicity, and nationality fail to currently provide a clear census. For instance, whilst \cite{83} and \cite{84} found that students from higher educational and socioeconomic classes were more likely to develop problematic cell phone usage habits, \cite{56} and \cite{57} found the converse to be true. Thus whilst questions of ethnic and socioeconomic differences in phone usage, including phone walking, and technology addiction are interesting and important to study further, it is difficult to formulate any precise hypotheses for phone walking from the existing literature.

From a different perspective, population subdivisions by age groups leads to striking differences in the rate of mobile phone addictions and as such provides an excellent source for proposing hypotheses for phone walking. Firstly, as might be anticipated, ownership rates of mobile phones, and in particular current generation smart phones, is lower in the elderly. However, there is evidence to suggest that the gap in smart phone ownership between the young and elderly is narrowing  [\cite{age3}]. Another relevant trend is that older phone users are typically more reluctant to embrace the full range of functionality of smart phones, often opting to use their devices mainly for calls and text messages [\cite{age1}]. These two factors taken together likely imply that one would typically expect to observe a higher rate in phone walkers in younger age ranges (for instance 20-45 years). 

%%%%%%
\newpage
%%%%%%
It is also interesting that the rate of phone usage has been observed to increase with age during the teenage years  \cite{age2}, and thus it is not immediately clear which age bracket would exhibit the peak rate of phone walking behaviour. Studies have found that 60\% of college students in the United States would self-declare as cell phone addicts [\cite{Roberts}] and in the 18-24 year age bracket 44\% have fallen asleep holding their phone in their hand. Moreover, whilst, 36\% of the entire  adult population self-report that they check their phones constantly, this rises to 54\% in young adults [\cite{BOA}] . Thus, drawing on the above, our hypothesis to be tested in future work, is that phone walking behaviour roughly mirrors the rates of constant checking observed in [\cite{BOA}] with respect to age, with individuals of ages 18-24 expected to be phone walkers roughly 50\% more frequently than the general adult population. Furthermore, we hypothesize that the rate of phone walker increases with age among teenagers, and peaks during the 18-24 age bracket.  It will be interesting to test these hypotheses in future studies.

 In summary the findings presented here clearly identify trends in how individuals interact with their inactive mobile devices in the overall adult population and note important divisions due to the gender composition of groups. In future work we intend to return to this topic to examine the variability in the rate of phone walking when the population is divided in manners other than gender. It will be interesting to see how ethnicity, class, age, and other factors influence the occurrence of phone walking and the use of mobile devices more generally.

   \section*{Ethic Declaration}
There are no ethical implications for this study and it is understood that the work undertaken does not classify as humans subjects research as defined by either the DHHS [45 CFR 46.102(f)] or FDA [21 CFR 50.3(c), 21 CFR 56.102(c)]. Only passive observations were conducted, which involved neither interventions or interactions. No private or personally identifiable information was observed or recorded, and persons of age under 21 years were not included in the sampling.

\end{document}